\documentclass[aps, prb, preprintnumbers, amsmath, amssymb, superscriptaddress, twocolumn, floatfix]{revtex4-1}
\usepackage{graphicx, color, dcolumn, natbib, bm, amssymb, amsmath, soul}
\usepackage[table,  dvipsnames]{xcolor}
\graphicspath{{fig/}}
\usepackage[final=true]{hyperref}
\begin{document}
\title{Anisotropic fractal magnetic domain pattern in bulk Mn$_{\text{1.4}}$PtSn}
\author{A. S. Sukhanov}
\affiliation{Max Planck Institute for Chemical Physics of Solids, D-01187 Dresden, Germany}
\affiliation{Institut f{\"u}r Festk{\"o}rper- und Materialphysik, Technische Universit{\"a}t Dresden, D-01069 Dresden, Germany}

\author{B. E. Zuniga Cespedes}
\affiliation{Max Planck Institute for Chemical Physics of Solids, D-01187 Dresden, Germany}
\affiliation{Institute of Applied Physics, Technische Universit{\"a}t Dresden, D-01069 Dresden, Germany}

\author{P. Vir}
\affiliation{Max Planck Institute for Chemical Physics of Solids, D-01187 Dresden, Germany}

\author{A. S. Cameron}
\affiliation{Institut f{\"u}r Festk{\"o}rper- und Materialphysik, Technische Universit{\"a}t Dresden, D-01069 Dresden, Germany}

\author{A.~Heinemann}
\affiliation{German Engineering Materials Science Centre (GEMS) at Heinz Maier-Leibnitz Zentrum (MLZ), Helmholtz-Zentrum Geesthacht GmbH, D-85747 Garching, Germany}

\author{N.~Martin}
\affiliation{Universit\'e Paris-Saclay, CEA, CNRS, Laboratoire L\'eon Brillouin,  CEA Saclay 91191 Gif-sur-Yvette, France}

\author{G. Chaboussant}
\affiliation{Universit\'e Paris-Saclay, CEA, CNRS, Laboratoire L\'eon Brillouin,  CEA Saclay 91191 Gif-sur-Yvette, France}

\author{V. Kumar}
\affiliation{Physik-Department, Technische Universit\"{a}t M\"{u}nchen, D-85748 Garching, Germany}

\author{P. Milde}
\affiliation{Institute of Applied Physics, Technische Universit{\"a}t Dresden, D-01069 Dresden, Germany}

\author{L. M. Eng}
\affiliation{Institute of Applied Physics, Technische Universit{\"a}t Dresden, D-01069 Dresden, Germany}
\affiliation{Dresden-W\"{u}rzburg Cluster of Excellence on Complexity and Topology in Quantum Matter (ct.qmat), TU Dresden, 01062 Dresden, Germany}

\author{C. Felser}
\affiliation{Max Planck Institute for Chemical Physics of Solids, D-01187 Dresden, Germany}
\affiliation{Dresden-W\"{u}rzburg Cluster of Excellence on Complexity and Topology in Quantum Matter (ct.qmat), TU Dresden, 01062 Dresden, Germany}

\author{D. S. Inosov}
\affiliation{Institut f{\"u}r Festk{\"o}rper- und Materialphysik, Technische Universit{\"a}t Dresden, D-01069 Dresden, Germany}
\affiliation{Dresden-W\"{u}rzburg Cluster of Excellence on Complexity and Topology in Quantum Matter (ct.qmat), TU Dresden, 01062 Dresden, Germany}

\begin{abstract}

The tetragonal compound Mn$_{1.4}$PtSn with the $D_{2d}$ symmetry recently attracted 
attention as the first known material that hosts magnetic antiskyrmions,
which differ from the so far known skyrmions by their internal structure.
The latter have been found in a number of magnets with the chiral crystal structure.
In previous works, the existence of antiskyrmions in Mn$_{1.4}$PtSn was unambiguously demonstrated in real space by means of Lorentz transmission electron microscopy on thin-plate samples ($\sim$100~nm thick).
In the present study, we used small-angle neutron scattering and magnetic force microscopy to perform reciprocal- and real-space imaging of the magnetic texture of bulk Mn$_{\text{1.4}}$PtSn single-crystals at different temperatures and in applied magnetic field.
We found that the magnetic texture in the bulk differs significantly from that of thin-plate samples.
Instead of spin helices or an antiskyrmion lattice, we observe an anisotropic fractal magnetic pattern of closure domains in zero field above the spin-reorientation transition temperature, which transforms into a set of bubble domains in high field.
Below the spin-reorientation transition temperature the strong in-plane anisotropy as well as the fractal self-affinity in zero field is gradually lost, while the formation of bubble domains in high field remains robust.
The results of our study highlight the importance of dipole-dipole interactions in thin-plate samples for the stabilization of antiskyrmions and identify criteria which should guide the search for potential (anti)skyrmion host materials. 
Moreover, they provide consistent interpretations of the previously reported magnetotransport anomalies of the bulk crystals.

\end{abstract}
 
\maketitle

\section{Introduction}

Solids exhibiting topological properties are promising for future applications, in particular for spintronics. In the case of magnetic materials with the ferromagnetic spin-spin exchange coupling, a presence of the antisymmetric Dzyaloshinskii-Moriya interaction (DMI) may twist the otherwise homogeneous collinear spin texture into a two-dimensional lattice of densely-packed nm-sized whirls. Each of these whirls is formed by a spatial distribution of the regularly canted magnetic moments that wrap a whole unit sphere, if mapped out onto it. The mutual noncoplanar orientation of the neighboring spins can be described by the topological charge (or the skyrmion winding number) $N_{\text{sk}}$ that takes values $\pm 1$ and differentiates two types of the topologically-protected magnetic structures -- skyrmions and antiskyrmions~\cite{Koshibae,Huang2017,Camosi,Hoffmann,Zhang,Nagaosa}.

Whilst there is an increasing number of the discovered skyrmion-hosting compounds, such as B20-type chiral magnets (e.g. MnSi, FeGe, Fe$_{\text{x}}$Co$_{\text{1-x}}$Si)~ \cite{Muehlbauer,Kindervater,Yu_2,Wilhelm,Milde2013,Moskvin}, $\beta$-Mn-type Co-Zn-Mn alloys~\cite{Tokunaga,Karube,Ukleev,Karube2,Karube3}, Cu$_2$OSeO$_3$~\cite{Seki12,Adams, Zhang2016, Milde2016, Stefancic,Sukhanov}, or lacunar spinels (e.g. GaV$_{\text{4}}$S$_{\text{8}}$, GaV$_{\text{4}}$Se$_{\text{8}}$)~\cite{Kezsmarki, Fujima, Bordacs}, which support Bloch-type or N\'eel-type skyrmions, respectively, antiskyrmions were observed only in thin-plates of Mn$_{\text{1.4}}$Pt(Pd)Sn~\cite{Nayak,Saha,Jena,Peng} and Mn$_2$Rh$_{0.95}$Ir$_{0.05}$Sn~\cite{Jena2} up to date. In contrast to the skyrmion materials with cubic ($P2_13$ or $P4_132$ space groups) or rhombohedral $C_{3v}$ crystal structures (space group $R3m$), tetragonal Mn$_{\text{1.4}}$PtSn belongs to the $D_{2d}$ symmetry class (space group $I\overline{4}2d$), which is a prerequisite of antiskyrmions~\cite{Hoffmann,Nayak,Meshcheriakova ,Bogdanov }. In agreement with the symmetry-based theoretical predictions, the first Lorentz transmission microscopy (LTEM) measurements of Mn$_{\text{1.4}}$PtSn demonstrated a nucleation of the triangular lattice of magnetic antiskyrmions in a magnetic field of $\sim$0.2~T, applied perpendicular to the surface of a thin lamella sample and parallel to the [001] crystallographic direction (the tetragonal $c$-axis). The antiskyrmions are $\sim$200~nm in diameter and were observed in a wide temperature range below $T_{\text C}$ of $\sim$400~K down to $T \approx 150$~K~\cite{Nayak,Saha}.

Subsequent LTEM experiments revealed that the antiskyrmions in Mn$_{\text{1.4}}$PtSn can also arrange in a square lattice in some particular temperature and field regions of the phase diagram, which may be affected by the sample thickness~\cite{Jena,Peng}. Moreover, elliptically-distorted skyrmions of both handedness and the non-topological bubble lattice  were shown to appear when a symmetry-breaking in-plane magnetic field is applied in a combination with the out-of-plane field~\cite{Jena,Peng}. This makes Mn$_{\text{1.4}}$PtSn a unique compound hosting a rich variety of controllable topological magnetic objects.

Because the previous studies on thin lamellae of Mn$_{\text{1.4}}$PtSn pointed out that the sample geometry and the sample preparation process can alter some aspects of the material properties~\cite{Jena,Peng}, it is essential to characterize the magnetic structure of the compound in the bulk.
Due to a very high penetration depth of the neutron radiation, neutron scattering techniques allow investigations of the magnetic structure of bulk samples.
Taking into account the long-periodic modulations of the magnetic texture in Mn$_{\text{1.4}}$PtSn, small-angle neutron scattering (SANS) is a suitable probe, which enables reciprocal-lattice imaging of magnetic structures with periods ranging from $\sim$2 to $\sim$400~nm~\cite{Muehlbauer2}.
For real-space investigations of the magnetic texture of bulk samples, mostly surface-sensitive techniques are available.
Magnetic force microscopy (MFM) proved to be a valuable tool when studying complex spin textures such as helices and skyrmions as well as complex domain patterns on length scales between $\sim$20~nm to $\sim$100~$\mu$m and can be applied at various temperatures as well as in external magnetic fields \cite{Milde2013, Milde2016, Kezsmarki, Hubert}.

In the present study, we employ both SANS and MFM to resolve the nm-scale magnetic texture of Mn$_{\text{1.4}}$PtSn in the bulk single-crystalline form and observe how it changes when the sample temperature and the applied magnetic field are varied. We discuss the characteristic features of the obtained reciprocal-space and real-space patterns and demonstrate that the magnetic structure of the bulk Mn$_{\text{1.4}}$PtSn differs dramatically from the previously reported LTEM observations obtained using samples in thin-plate geometry.
Instead of helices or an antiskyrmion lattice, we observe an anisotropic fractal magnetic pattern of closure domains in zero field above the spin-reorientation transition temperature $T_{\text{SR}}$, with characteristic hints for the DMI inherent to the $D_{2d}$ symmetry of the crystal, which transforms into a set of bubble domains in high field.
Below $T_{\text{SR}}$ the strong in-plane anisotropy as well as the fractal self-affinity in zero field are gradually lost, while the formation of bubble domains in high field remains robust.
The results of our study highlight the importance of dipole-dipole interactions in thin-plate samples for the stabilization of antiskyrmions and identify search criteria for potential (anti)skyrmion host materials.

The paper is organized as follows.
After a description of the experimental details, we first discuss the experimental results obtained in zero field and above $T_{\text{SR}}$.
Then, we describe the field dependency above $T_{\text{SR}}$.
Finally, we discuss the temperature dependence.

\section{Experimental details}

Our SANS measurements were performed at the instruments SANS-1 (FRM-II, Garching, Germany) and PA20 (LLB-Orph\'{e}e, CEA Saclay, France)~\cite{PA20}. In both experiments, we used the same sample that consisted of 17 crystals coaligned together with a relative misalignment not worse then 3$^{\circ}$, mounted on an aluminum plate holder. The crystals were coaligned to increase the total volume of the sample and the resulting experimental signal-to-noise ratio (see the supplemental materials~\cite{Supp} for the photograph of the sample). All the crystals were grown by the self-flux method, as described in~\cite{Vir1,Vir2}. The high quality of the crystals was confirmed by means of magnetic susceptibility, resistivity, and x-ray diffraction measurements. Their stoichiometry was examined by energy-dispersive x-ray spectroscopy.

MFM measurements were performed in two instruments. For room-temperature measurements without external fields we used the Park Systems NX10~\cite{Park} with PPP-MFMR probes from Nanosensors~\cite{Nanosensors} at lift heights between 100 and 150~nm. Low-temperature measurements with external field were performed in an Omicron cryogenic ultra-high vacuum STM/AFM instrument~\cite{Omicron} using the RHK~R9s electronics~\cite{RHK} for scanning and data acquisition.
We employed PPP-QMFMR probes from Nanosensors driven at mechanical oscillation amplitudes $A \approx 20$~nm at lift heights between 400 and 800~nm.
All data analysis was performed with the Gwyddion~\cite{Gwyddion} software.
Two samples have been investigated. Sample A is a single crystal of Mn$_{\text{1.4}}$PtSn, whose native surface was gently polished with a focussed ion beam using Xe ions at currents below 10~nA. After the polishing we checked with MFM, that the resulting amorphous surface layer did not alter the domain pattern. Sample B is a polycrystalline sample of Mn$_{\text{1.4}}$Pt$_{\text{0.9}}$Pd$_{\text{0.1}}$Sn that was carefully polished. Measurements of sample A are presented in the paper, while measurements on sample B can be found in the supplemental material.

\section{Results}

\subsection{Magnetic texture at $T>T_{\text{SR}}$}

\begin{figure}
\includegraphics[width=0.99\linewidth]{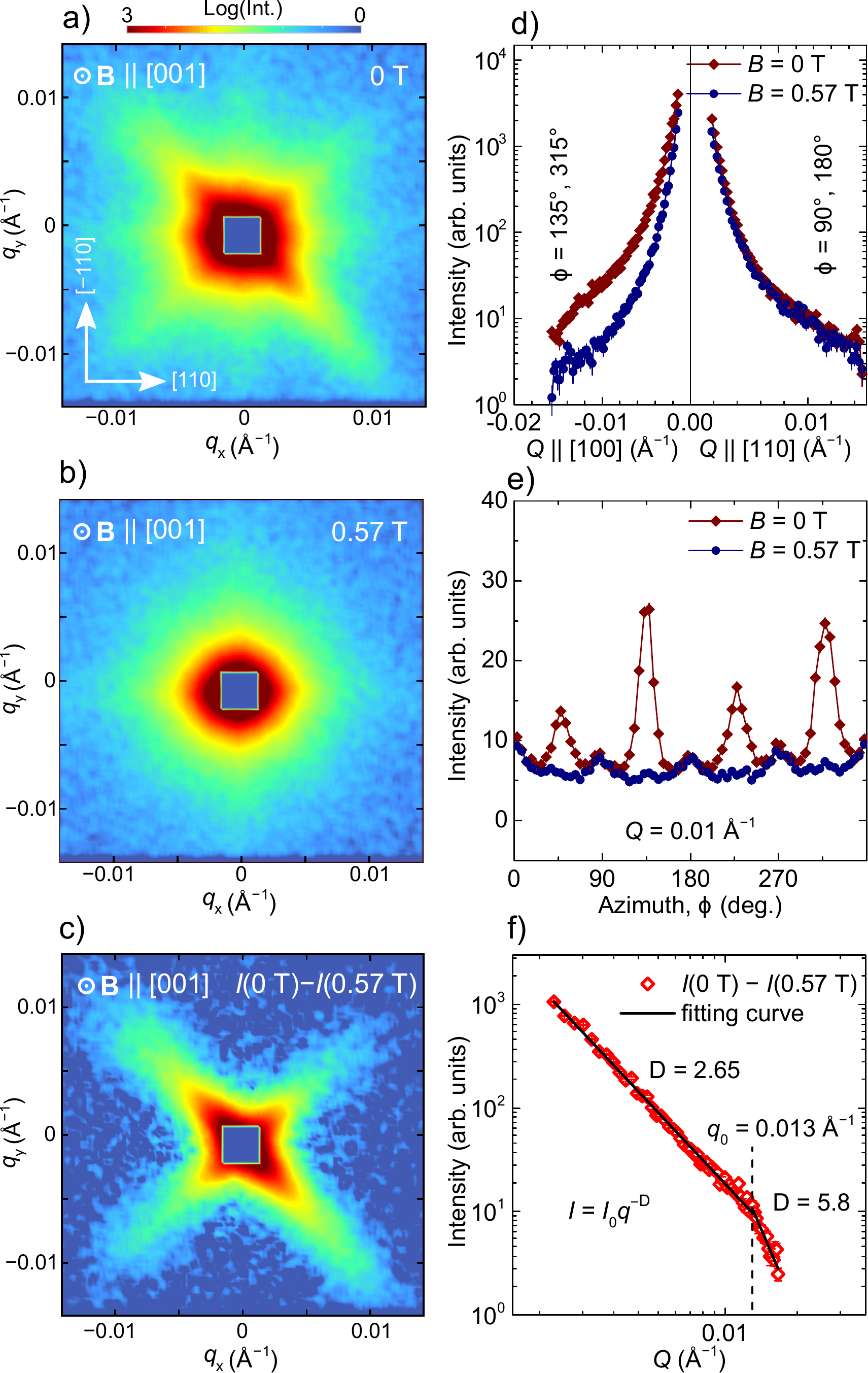}\vspace{3pt}
        \caption{(color online). The SANS patterns of Mn$_{\text{1.4}}$PtSn at 250~K. (a) The SANS pattern at zero field. The sample orientation is depicted by white arrows. (b) The scattering in the field-saturated state at 0.57~T. (c) The result of the subtraction of the SANS map recorded at 0.57~T from the map at 0~T. (d) The comparison of the intensity profiles for the momenta along [100] and [110]. (e) The intensity profiles as a function of the azimuthal angle $\phi$. (f) The radial intensity profile of the background-subtracted data (symbols) and the fit by the power function (solid line).}
        \label{ris:fig1}
\end{figure}

Fig.~\ref{ris:fig1}(a) shows a SANS pattern collected at the sample temperature $T = 250$~K and zero magnetic field (no prior field history). The sample was oriented with its tetragonal $c$-axis parallel to the incident neutron beam. In this scattering geometry, the reciprocal $(HK0)$ plane is imaged at the position-sensitive detector. The in-plane orientation of the sample corresponds to the momentum component $q_x$ aligned with the [110] direction. Thus, the scattering pattern in Fig.~\ref{ris:fig1}(a) represents spin-texture modulations in the $ab$-plane of Mn$_{\text{1.4}}$PtSn. As can be seen, there is a clear scattering intensity distribution that covers almost the whole imaged reciprocal space in the momentum range 0.002~\AA$^{-1} < Q < 0.013$~\AA$^{-1}$. The scattering is diffuse and does not have sharp features, such as Bragg peaks one would expect in SANS of helimagnets~\cite{Muehlbauer,Kindervater,Moskvin,Tokunaga,Adams,Sukhanov}. This is in strong contrast to the LTEM observations of the helical spin structure in the thin lamellae samples, including at $T = 250$~K and $B = 0$~T~\cite{Nayak,Saha,Jena,Peng}, and implies that the magnetic structure of the bulk Mn$_{\text{1.4}}$PtSn is not a spin helix, as was previously anticipated. The observed SANS pattern is diffuse yet  strongly anisotropic and can be viewed as eight streaks/stripes of intensity, which can be considered as a sum of two four-pointed stars/crosses, one of which is higher in intensity and points to the $\langle 100 \rangle$ directions, whereas the other is weaker and oriented with respect to $\langle 110 \rangle$.

The origin of this scattering can be understood when a saturating magnetic field is applied parallel to the $c$-axis (i.e. along the neutron beam). Fig.~\ref{ris:fig1}(b) depicts the SANS pattern at $B = 0.57$~T, which is well above the saturation field of $\sim$0.5~T~\cite{Vir1}. As one can see, the stripes of intensity along $\langle 100 \rangle$ disappeared but the weaker stripes along $\langle 110 \rangle$ remained unchanged, which implies that only the former are of magnetic origin. Hence, the pattern in Fig.~\ref{ris:fig1}(b) can be used as background, and we subtracted the intensity $I(\textbf{Q}, 0.57~\mathrm{T})$ from the pattern $I(\textbf{Q}, 0~\mathrm{T})$. 
The resulting pattern is shown in Fig.~\ref{ris:fig1}(c), which represents the pure magnetic intensity.
To highlight the difference between the zero-field and the field-saturated states, intensity profiles are plotted for the momentum directions $(H00)$ and $(HH0)$ in Fig.~\ref{ris:fig1}(d). Fig.~\ref{ris:fig1}(e) demonstrates the anisotropy of the intensity distribution cut at $Q = 0.01$~\AA$^{-1}$ ~as a function of the in-plane (azimuthal) angle $\phi$. The perpendicular cut through each stripe is a peak with a base width of $\sim$45$^\circ$.

The observation of diffuse anisotropic scattering suggests that Mn$_{\text{1.4}}$PtSn develops a magnetic texture of rectangular-shaped domains with domain walls oriented strictly perpendicular to the crystallographic [100] and [010] axes. The well-defined orientation of the magnetic domains follows from the cross-shaped scattering within the reciprocal $(HK0)$ plane. The domains, however, do not feature any regularity either in their sizes or the domain-wall spacing, which can be concluded from the smooth radial profile of the diffuse scattering. 

For further analysis, the radial profile along the stripe of the background-subtracted SANS pattern taken at $T=250$~K and $B=0$~T was plotted in Fig.~\ref{ris:fig1}(f) on a log-log scale. 
$I(q)$ obeys a power-function trend  $I \propto q^{-D}$ with two different exponents below and above the crossover momentum $q_0 = 0.013$~\AA$^{-1}$, where the slope  changes. The fitting yields $D = 2.65$ for momenta below $q_0$ down to the lowest accessible momentum of $\sim$0.002~\AA$^{-1}$, which is a signature of the scattering from fractal objects~\cite{Schmidt,Mildner,Martin,Iashina}.
This implies a complex intertwined arrangement of the domain walls of the rectangular domain pattern.
The momentum $q_0$ then determines the lowest real-space scale down to which the fractal self-affinity holds, which is here $\sim$48~nm. The upper limit of the fractal structure cannot be reached within the accessible $q$ range and lies above 315~nm.
Above $q_0$ the exponent $D$ changes to $D=5.8$, which can be attributed to scattering from density profiles without sharp ($D = 4$) contrast (either due to roughness along the profile or a smoothness of the profile)~\cite{Schmidt,Mildner}. 
A similar analysis was applied in a SANS study of Nd$_{\text{2}}$Fe$_{\text{14}}$B \cite{Kreyssig}, where anisotropic diffuse scattering described by $D = 3.7$ ($D = 3.1$) was observed below (above) the spin-reorientation transition.\\

\begin{figure}
	\includegraphics[width=0.99\linewidth]{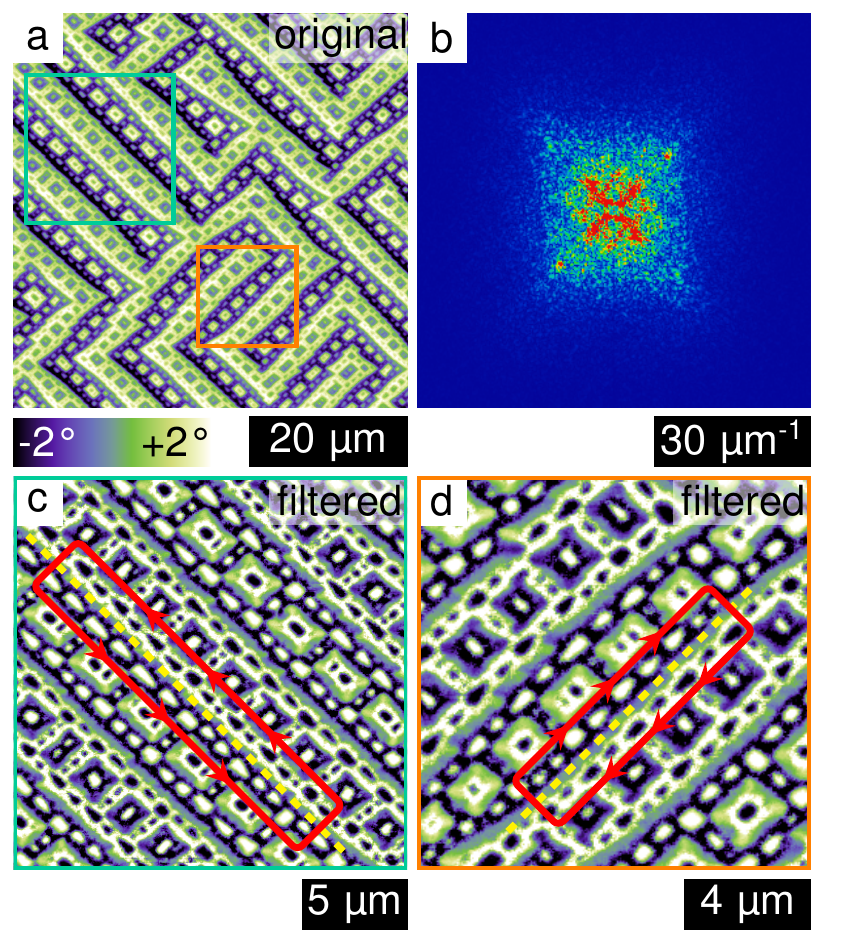}
	\caption{(color online). The magnetic domain pattern of Mn$_{\text{1.4}}$PtSn at room temperature. (a) MFM phase image showing the domain pattern in the $ab$-plane at zero field, and (b) its Fast Fourier transform. (c) and (d) depict the zoomed area marked in (a) by the turquoise and orange squares, respectively. They are filtered for enhanced contrast of the smaller domains. The red loops highlight the chiral sense of the pattern of triangular shaped nested domains at the domain walls (yellow dashed line) of the lamellar stripe domains.}
	\label{ris:fig1a}
\end{figure}

For comparison, we show a MFM measurement on the $ab$-plane of the Mn$_{\text{1.4}}$PtSn single-crystal (sample A) at room temperature in Fig.~\ref{ris:fig1a}(a). The domain pattern is highly reminiscent of the fractal magnetic closure domains observed in the low-temperature easy-cone anisotropy phase of Nd$_{\text{2}}$Fe$_{\text{14}}$B~\cite{Pastushenkov1997, Kreyssig, Pastushenkov2010}.
It consist of lamellar domains with smaller nested closure domains arranged in Sierpinski carpets.
The width of the lamellar domains sets the upper boundary of $\sim$3~$\mu$m, which is one order of magnitude higher than the upper fractal scale accessed in SANS.
The in-plane orientation of the domain walls is highly anisotropic with pinning to two perpendicular directions within the $ab$-plane.
In the Fourier transform [see Fig.~\ref{ris:fig1a}(b)], the same cruciform pattern as in SANS is visible.
Hence, we can safely assume, that our SANS and MFM results describe the same anisotropic fractal domain pattern.
Such domain patterns arise from a competition of uniaxial ferromagnetic exchange interaction favoring collinear domains in the easy axis without domain walls and dipolar interactions at the surfaces of the material adding stray-field energy. The latter is minimized by nucleation of closure domains at the surface at the expense of additional domain walls, whose orientation is defined by the anisotropy within the $ab$-plane. Thus, the nested domains are present in a region below the surface and form a fractal tree structure along the $c$-axis, also known as branching domains~\cite{Hubert}.
Additional measurements on the polycrystalline sample B confirm, that the fractal pattern indeed belongs to such closure domains. Details can be found in the supplemental material.

Moreover, we find characteristic hints for the DMI in the material.
As already mentioned, lamellar stripe domains appear in two orientations within the $ab$-plane.
Examples are shown in Figs.~\ref{ris:fig1a}(c,d).
Note, that these two images have been filtered for enhanced contrast of the smaller domains.
The nested domains within these stripes partly appear with an arrowhead shape.
The direction defined by the arrows along the stripe domain walls (highlighted by dotted yellow lines) defines a certain chirality, which is solely set by the in-plane orientation of the domain wall, as schematically shown in Figs.~\ref{ris:fig1a}(c,d) by the red loops.
This reflects the $D_{2d}$ symmetry of the crystal, namely, in order to transform from one loop to the other, one has to apply the combination of a $90^\circ$ rotation and an inversion, like for the chirality of helices and the non-topological bubbles observed by LTEM in thin-plate samples~\cite{Peng}.
This very peculiar feature of the domain pattern is so far unique for Mn$_{\text{1.4}}$PtSn and to the best of our knowledge has not been reported before for any other fractal magnetic domain pattern.
It may be related to the Dzyaloshinski-Moriya interaction, which is responsible for the existence of antiskyrmions in the material in the first place~\cite{Nayak}.

\subsection{Domain structure in applied field}

\begin{figure}
	\includegraphics[width=0.99\linewidth]{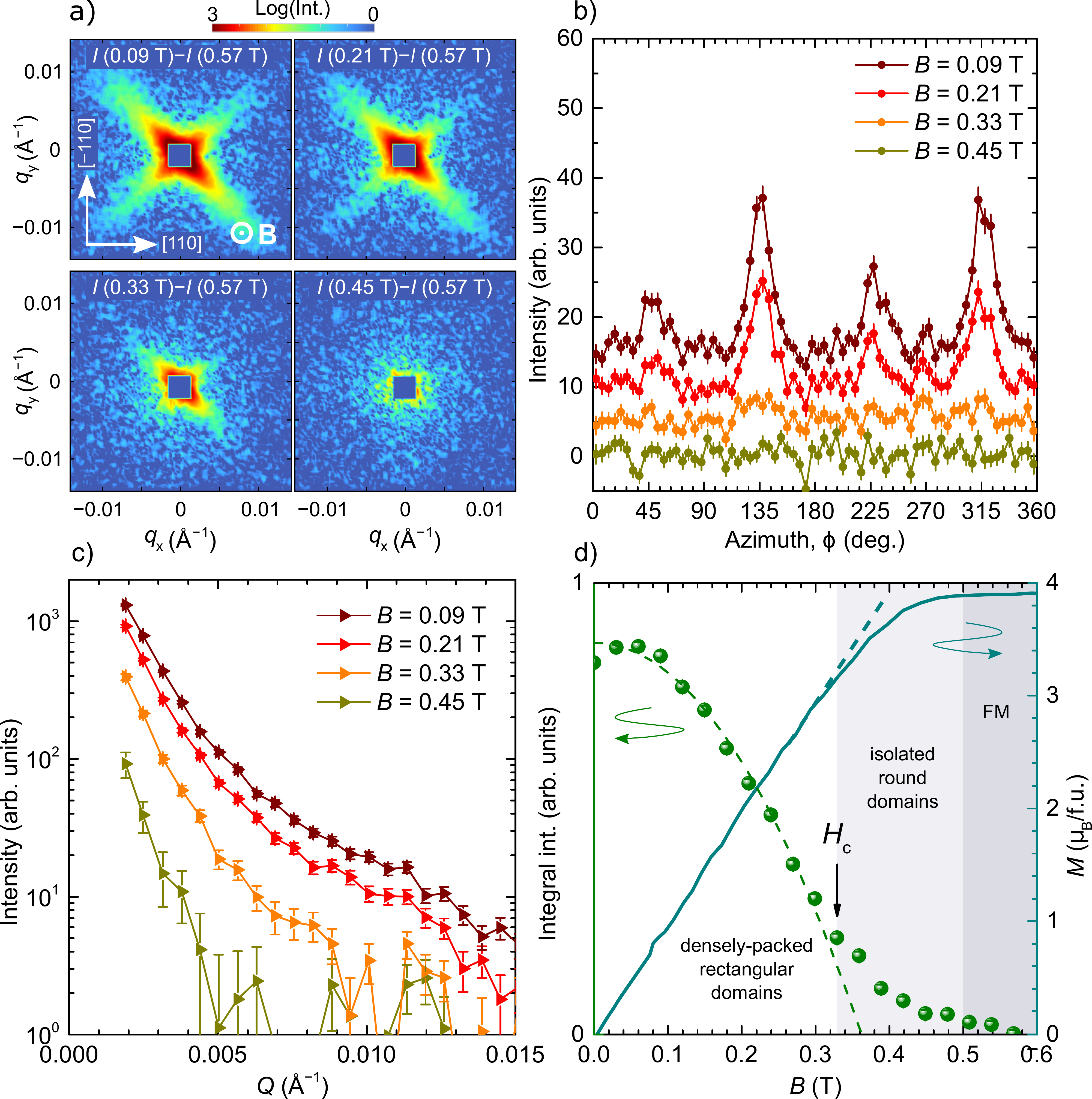}\vspace{3pt}
	\caption{(color online). Magnetic field dependence of the SANS signal at $T = 250$~K. (a) A series of SANS patterns (after background subtraction) collected in different magnetic fields applied along the $c$-axis. (b) The azimuthal profiles of the scattering extracted from each pattern. (c) The radial intensity profiles at the same fields. (d) The integral intensity of the magnetic scattering as a function of the applied field (symbols) and the magnetization $M(H)$ curve measured at the same temperature and field orientation. The dashed lines are the approximations by a parabolic and a linear functions (see the text). The black arrow marks the crossover between the two field-polarizing regimes.}
	\label{ris:fig2}
\end{figure}

Next, we discuss the magnetic-field response of the magnetic texture of Mn$_{\text{1.4}}$PtSn.
Fig.~\ref{ris:fig2}(a) shows a series of SANS patterns collected at different field magnitudes applied parallel to the $c$-axis at $T = 250$~K.
For all patterns, the $B = 0.57$~T scattering (the fully-polarized state) was subtracted as background in analogy to Fig.~\ref{ris:fig1}(c).
The azimuthal intensity profiles were extracted from each SANS pattern and plotted in Fig.~\ref{ris:fig2}(b). The pattern of $B = 0.09$~T looks very similar to the SANS pattern at zero field, namely, it has the same anisotropic cruciform scattering distribution with the same intensity. The pattern recorded at $B = 0.21$~T retains the anisotropy with approximately two times lower overall intensity. The SANS pattern at $B = 0.33$~T demonstrates very low intensity, which is seen only in the vicinity of the center $Q \approx 0$. Nevertheless, the characteristic cross shape remains well distinguishable in the pattern. At a higher field of 0.45~T, only a very weak isotropic scattering is observed at small momenta in the vicinity of the direct beam,
which indicates a transformation from rectangular towards isotropic domains.
Not only is the symmetry of the diffuse scattering preserved in increased magnetic field up to $B = 0.33$~T, but also the radial $I(Q)$ profiles approximately retain the initial slope, as evidenced in Fig.~\ref{ris:fig2}(c), where the profiles are shown for the same set of fields.\\

There are two possible scenarios of the magnetic-structure polarization process that would cause the observed smooth decrease in the intensity of the anisotropic diffuse scattering. The first one implies that in the applied magnetic field the rectangular-domain texture is gradually dissolved into the homogeneous ferromagnetic background. In other words, the sample breaks into coexisting domains of the fully-polarized state, growing in volume with an increasing field, and the volume occupied by the densely-packed small rectangular domains. In this case, the intensity of the SANS should decrease in accord with the modulated-texture volume, $I = I_0\left(1 - M_z/M_\mathrm{s}\right)$, where $M_z/M_\mathrm{s}$ is the normalized net magnetization, and $I_0$ is the intensity in zero field. In the opposite scenario, the modulated magnetic texture occupies the entire volume in finite applied fields, but the magnitude of the modulated component of the local magnetization (its in-plane projection) is reduced in favor of the homogeneous $M_z$ component. Since the SANS intensity $I \propto M_{x,y}(x,y)^2$, where $M_{x,y}(x,y)$ is the magnetization component modulated in the basal plane, the field-dependence of $I$ should read as $I = I_0\left[ 1 - \left( M_z/M_\mathrm{s} \right)^2 \right]$.

The integral intensities (integrated along the stripes) of the diffuse scattering were extracted from the SANS patterns measured in dependency of the field magnitude and plotted in Fig.~\ref{ris:fig2}(d) (symbols) along with the isothermal magnetization curve (solid line) obtained with a SQUID magnetometer. The magnetization demonstrates a linear dependence up to $B \approx 0.33$~T, where it reaches $\sim$3/4 of the saturated moment $M_\mathrm{s} = 3.9\mu_{\text B}$. Notably, the intensity $I(B)$ can be well approximated by a parabolic function in the same field range, which agrees with the second scenario. Above $B = 0.33$~T, the intensity starts deviating from the quadratic field dependence and switches to the $1 - M_z/M_\mathrm{s}$ behavior, as predicted by the first scenario, until it vanishes at the saturating field of $\sim$0.5~T. The field $B_{\text c} = 0.33$~T can be therefore denoted as the crossover point at which the partially-polarized rectangular domain texture is becoming diluted by the regions of the fully-polarized state or isotropic domains.\\

\begin{figure}
	\includegraphics[width=\columnwidth]{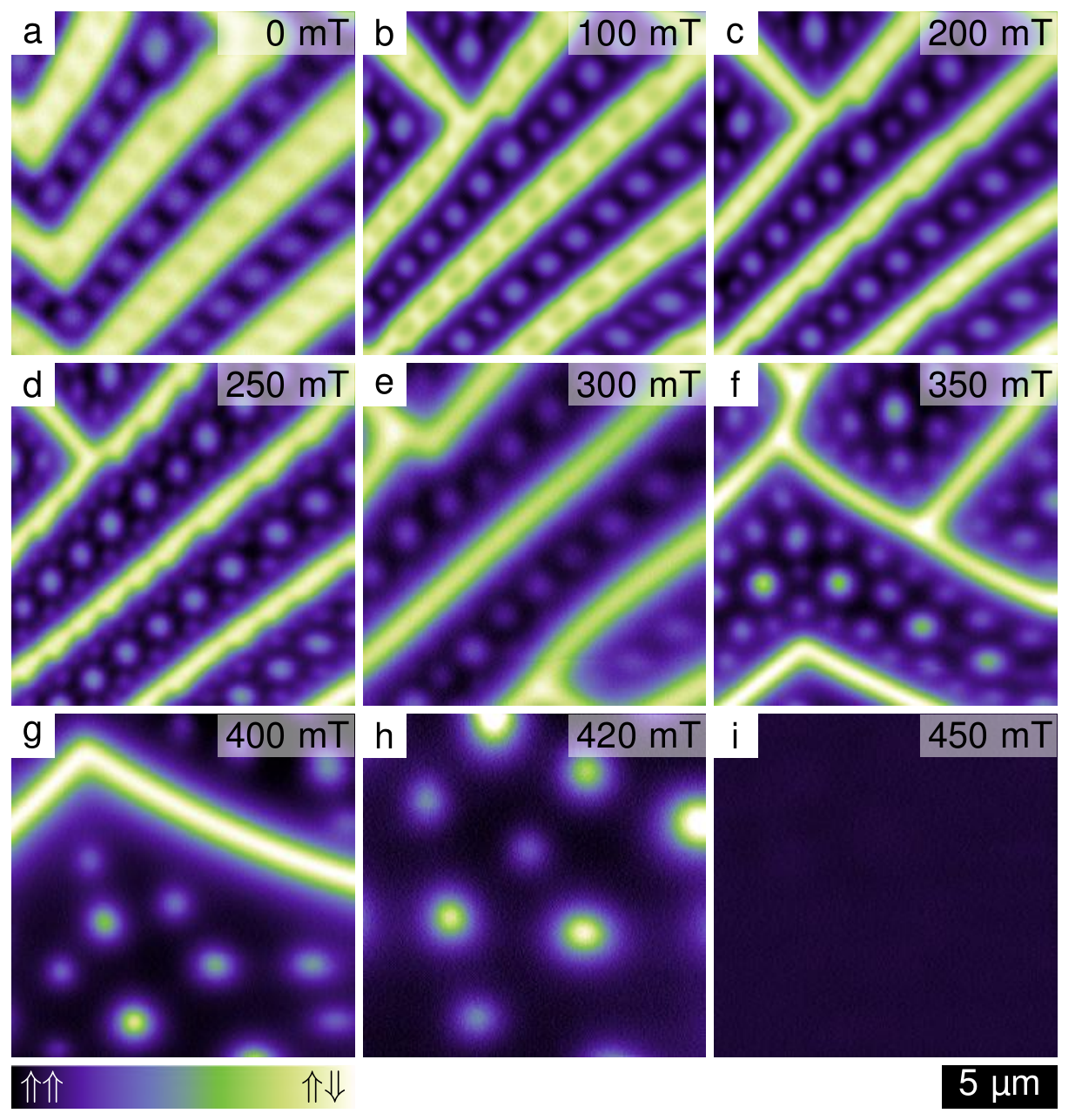}
	\caption{(color online). Magnetic field dependence of the domain pattern in MFM at $T=240$~K. Sample magnetization parallel or antiparallel to the external field is shown in purple or green, respectively. The frame size is equal for all images, the span of the color scale has been adapted individually.}
	\label{ris:fig2a}
\end{figure}

In Fig.~\ref{ris:fig2a}, we show a series of MFM measurements obtained at $T = 240$~K on the $ab$-plane of the Mn$_{\text{1.4}}$PtSn single-crystal with magnetic field applied parallel to the $c$-axis, i.e. perpendicular to the plane of view.
Areas with sample magnetization pointing parallel/antiparallel to the applied field are shown in purple/green, respectively. 
In the region between zero field and $B=100$~mT there are no qualitative changes in the domain pattern [see Figs.~\ref{ris:fig2a}(a,b)], which agrees with only a small change in SANS intensity in this field range.
With increasing field up to $B=200$~mT, shown in Fig.~\ref{ris:fig2a}(c), the nested domains within the antiparallel magnetized stripe domains (shown in green) disappear, yet the edges of those obey a sawtooth shape. In turn, in the stripe domains magnetized parallel to the field, the number of nested domains increases.
Between $B=250$ and 350~mT, the sawtooth shape of the domain walls is lost, and the nested domains appear with both a more rounded shape and  less ordering, shown in Figs.~\ref{ris:fig2a}(d) to \ref{ris:fig2a}(f).
For further increasing field magnitudes, the antiparallel magnetized domains start to disappear and only a few bubble domains or very large antiskyrmions are left over for $B > 400$~mT [see Fig.~\ref{ris:fig2a}(h)]. The last of those switch into the field-polarized state when $B = 450$~mT is reached, consistently with the isothermal magnetization curve taking into account the limited field of view of the MFM.
Overall, there is perfect agreement between the real-space images and the behaviour of the diffuse scattering pattern described before.

\subsection{Magnetic texture at $T<T_{\text{SR}}$}

\begin{figure}[b]
	\includegraphics[width=0.96\linewidth]{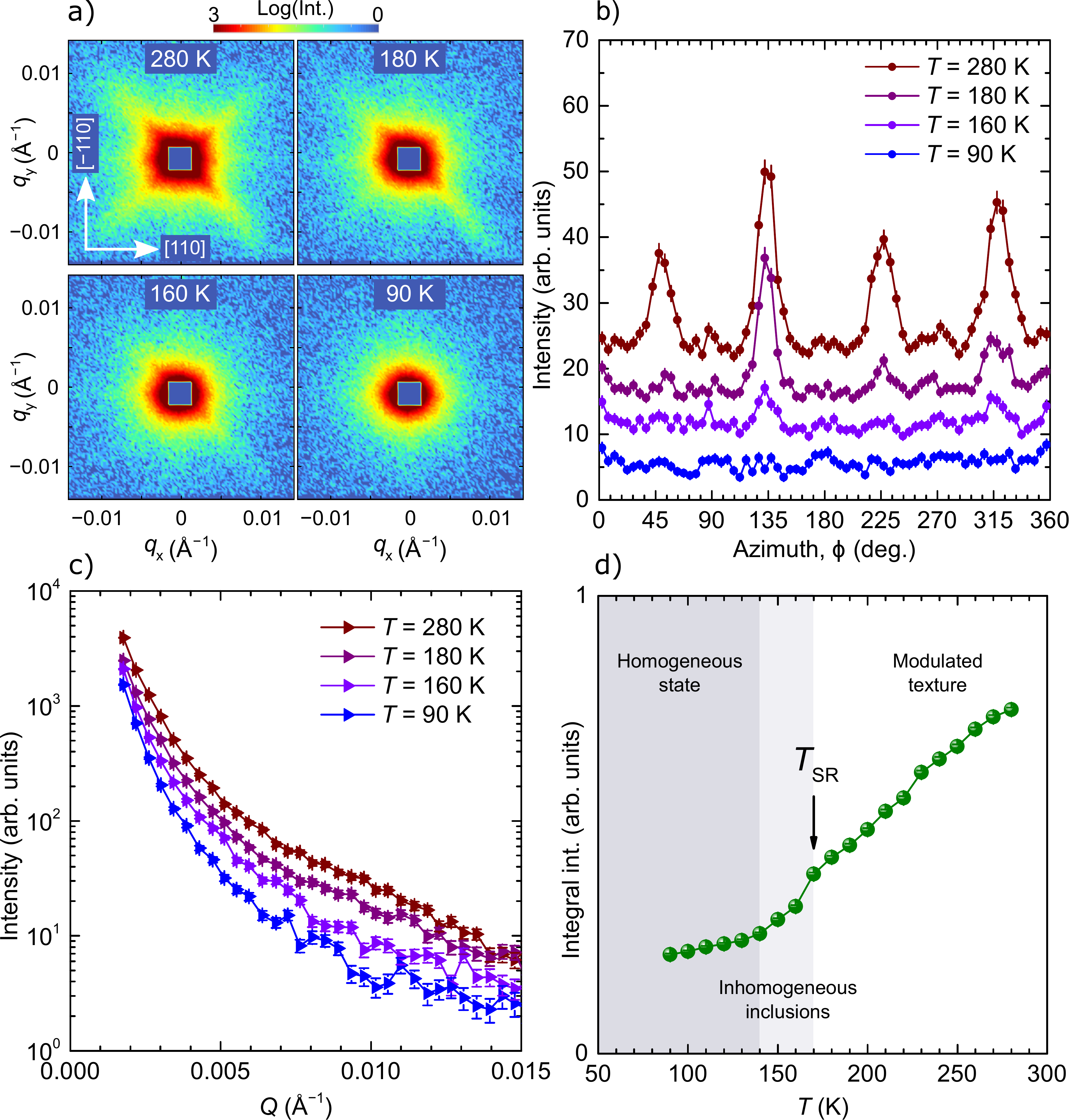}\vspace{3pt}
	\caption{(color online). Temperature dependence of the SANS signal at zero field. (a) A series of SANS patterns (no background subtraction) collected at different temperatures in zero-field cooling. (b) The azimuthal profiles of the scattering extracted from each patter. (c) The radial intensity profiles at the same temperatures. (d) The integral intensity of the magnetic scattering as a function of $T$. The black arrow marks the spin-reorientation transition.}
	\label{ris:fig3}
\end{figure}

Finally, it is important to understand how the magnetic texture changes with temperature. The crucial point is whether or not the rectangular domain structure persists at any $T < T_{\text C}$, including temperatures below the spin-reorientation (SR) transition $T_{\text{SR}} = 170$~K. The SR transition in Mn$_{\text{1.4}}$PtSn can be detected by a step in the temperature-dependent magnetic susceptibility and by a kink in the resistivity~\cite{Vir1,Vir2}. Powder neutron diffraction measurements showed that the local (on the scale of the unit cell) magnetic structure of Mn$_{\text{1.4}}$PtSn  is collinear ferromagnetic above $T_{\text{SR}}$ and becomes canted (noncollinear) ferromagnetic below $T_{\text{SR}}$~\cite{Vir1,Kumar}. The correlations between the local (within the unit cell) magnetic configuration and the topology of the large-scale magnetic texture seemed controversial. On the one hand, the magnetotransport measurements of bulk single crystals revealed a topological Hall effect (THE) below $T_{\text{SR}}$~\cite{Vir1}, which is widely associated with the skyrmion phase in other materials~\cite{Neubauer,Huang2012,Gallagher,Kanazawa,Yokouchi,Li,Franz} (antiskyrmions are expected to give rise to a THE similarly to skyrmions~\cite{Nagaosa}). On the other hand, the LTEM measurements~\cite{Nayak,Saha,Jena,Peng} demonstrated that the antiskyrmions nucleate only at $T > T_{\text{SR}}$, which suggests that the THE is related to the local noncollinear structure.

\begin{figure}[b]
	\includegraphics[width=0.96\linewidth]{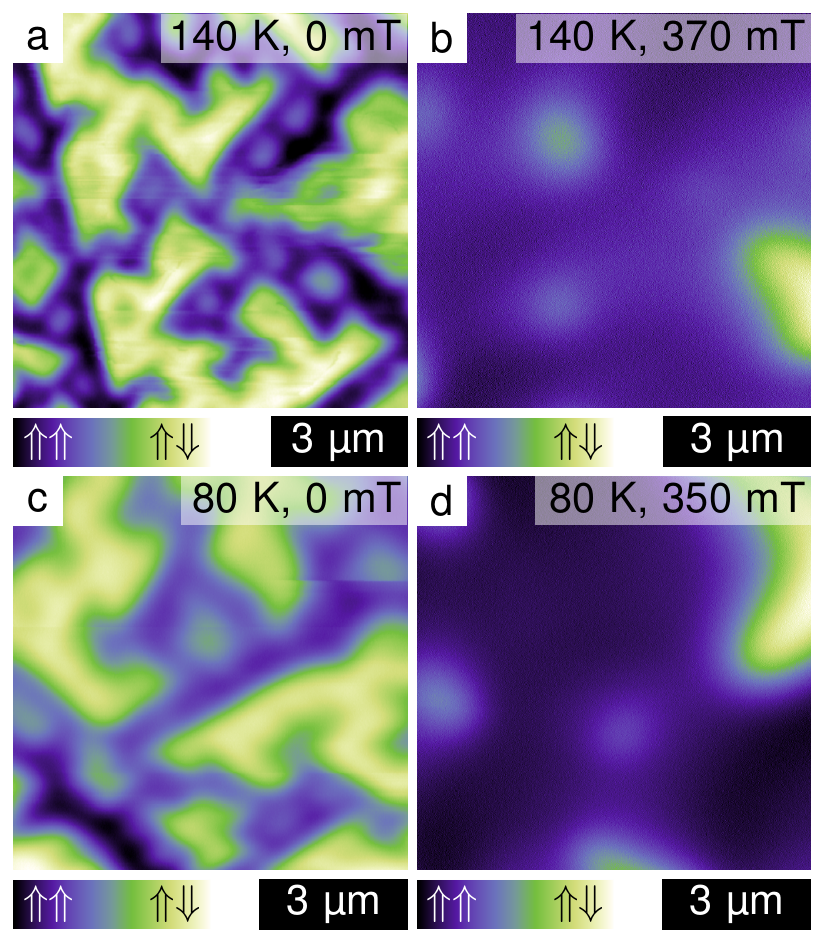}
	\caption{(color online).  Magnetic domain pattern in MFM below $T_{\text{SR}}$ at $T = 140$~K (a,b) and $T = 80$~K (c,d) both in zero field (a,c) and in high field (b,d). Sample magnetization parallel or antiparallel to the external field is shown in purple or green, respectively. The span of the color scale has been adapted individually.}
	\label{ris:fig3a}
\end{figure}

Again, we first describe the SANS measurements.
Fig.~\ref{ris:fig3}(a) shows SANS patterns (no background subtraction) collected after zero-field cooling to two temperatures above $T_{\text{SR}}$, namely $T = 280$ and 180~K, as well as to two temperatures below $T_{\text{SR}}$, namely $T = 160$~K and $T = 90$~K. The corresponding azimuthal profiles and the radial profiles of the intensity are plotted in Figs.~\ref{ris:fig3}(b) and \ref{ris:fig3}(c), respectively. The patterns at 280 and 180~K exhibit the same anisotropy. Despite the fact that the magnetic moment is reduced at higher temperatures due to thermal fluctuations (which reduces the intensity of magnetic scattering at elevated temperatures), the intensity at 280~K is significantly higher. At $T = 160$~K, which is just below $T_{\text{SR}}$, the stripes of intensity can be still distinguished but have much weaker intensity than at $T = 180$~K. Far below $T_{\text{SR}}$ at 90~K, the cross-shaped scattering disappears completely. The remaining diamond-shaped scattering is similar to the pattern at the field-polarized state at 250~K [Fig.~\ref{ris:fig1}(b)].  Finally, in Fig.~\ref{ris:fig3}(d), the scattering intensity of the $\langle 100 \rangle$ streaks is plotted as a function of temperature. It decreases upon cooling with a kink at $T = T_{\text{SR}}$. This shows that the fractal magnetic domain pattern is inherent to the high-temperature phase with the locally-collinear magnetic order. However, it does not transform to the homogeneous state immediately below the SR transition. Instead, traces of the anisotropic scattering are present over a 20--30~K wide region below $T_{\text{SR}}$, which correlates with the gradual change of the spin canting within the unit cell~\cite{Vir1,Kumar}.

The loss of the fractal self-affinity as well as the absence of lamellar stripe domains in zero field is apparent from the MFM measurements obtained below $T_{\text{SR}}$ both at $T = 140$ and 80~K, which we show in Figs.~\ref{ris:fig3a}(a,c) respectively.
Moreover, also the strong anisotropic pinning of the domain walls to two perpendicular directions within the $ab$-plane is lost.
Interestingly, in higher fields again round shaped domains appear before the field polarized state is reached [see Figs.~\ref{ris:fig3a}(b,d)]. In comparison to the reported LTEM measurements~\cite{Nayak,Saha,Jena,Peng}, where antiskyrmions were present only above $T_{\text{SR}}$, it is very likely that the domains here are (closure) bubble domains rather than antiskyrmions.

\section{Discussion and conclusions}

To conclude, we used a combination of SANS and MFM to study the bulk magnetic structure of Mn$_{\text{1.4}}$PtSn and showed that it differs drastically from what was previously reported from LTEM measurements of thin-plate samples. The bulk Mn$_{\text{1.4}}$PtSn does not support antiskyrmions or any other type of regular long-periodic single-$q$ or multi-$q$ structures, but develops ferromagnetic lamellar stripe domains combined with an anisotropic fractal surface domain pattern that has a characteristic scale with the lower boundary of $\sim$48~nm and the upper boundary $\sim$3~$\mu$m defined by the width of the lamellar stripe domains. This magnetic pattern gives rise to an anisotropic diffuse intensity distribution in SANS, and to the best of our knowledge was previously only observed in the low temperature phase of Nd$_{\text{2}}$Fe$_{\text{14}}$B. Yet, it differs from the latter in the presence of characteristic hints for the DMI inherent to the $D_{2d}$ symmetry of the crystal, which manifest in the orientation of arrowhead-shaped nested domains at the domain walls of the lamellar stripe domains.
Our measurements showed that the magnetic texture of bulk Mn$_{\text{1.4}}$PtSn is polarized by the applied magnetic field in a multi-step process. During the first step, the spins gradually cant towards the field direction keeping the overall domain pattern unaffected.
During the second step, the fractal domain pattern softens and the nested domains transform into an assembly of bubble domains.
Finally, in the third step, only the bubble domains persist and switch individually into the field polarized state.
The latter two steps resemble the metamagnetic phases of the dense antiskyrmion lattice and the isolated antiskyrmions in the ferromagnetic background, respectively, observed in the thin plates~\cite{Nayak,Saha,Jena,Peng}. 
Like the antiskyrmions in the thin-plate geometry, the fractal magnetic pattern of the bulk Mn$_{\text{1.4}}$PtSn appears for $T > T_{\text{SR}}$ and shows an enhanced stability at elevated temperatures, which further demonstrates the intimate connection between the magnetic structure in the bulk and in the thin plates.

To the best of our knowledge, Mn$_{\text{1.4}}$PtSn is therefore the first known material where the magnetic texture can be unambiguously tuned between ferromagnetic domains with an anisotropic fractal closure domain pattern in zero field or bubble domains in high field in bulk samples, and helices or antiskyrmions in the thin-plate geometry.
Our findings highlight the importance of the dipolar interaction, which was also realized to be essential to explain all the observations in recent LTEM experiments~\cite{Jena,Peng}.

Moreover, we may speculate that new materials may be (anti)skyrmion hosts in the thin-plate geometry even if they have not been identified as such in bulk experiments.
Possible candidates would be uniaxial materials with fractal closure domains in bulk and with $D_{2d}$ symmetry. Since only a non-centrosymmetric crystal symmetry is the necessary condition for the possible existence of (anti)skyrmions, the range of materials can be even larger.
Vice versa, like for Mn$_{\text{1.4}}$PtSn, materials that support (anti)skyrmions in the thin-plate geometry can show more complicated domain patterns in bulk. For example, this could be the case for Cr$_{\text{11}}$Ge$_{\text{19}}$, which hosts biskyrmions in thin-plate geometry~\cite{Takagi}.

When finalizing our manuscript, we became aware of a thickness-dependent study in the range up to 4~$\mu$m of the magnetic texture in  Mn$_{\text{1.4}}$PtSn~\cite{Ma}. The authors explained the crucial role of dipolar forces in the material by rigorous simulations, which fully agree with the conclusions drawn in our study.

\section*{Acknowledgments}

A.S.S. thanks S. E. Nikitin for many stimulating discussions. 
B.E.Z.C. thanks U. Burkhardt and S. Kostmann for electron backscatter diffraction measurements and polishing of the polycrystalline sample.
This project was funded by the German Research Foundation (DFG) under Grants No. FE~633/30-1, IN~209/9-1, EN~434/38-1, and MI~2004/3-1; as well as EN~434/40-1 and IN 209/7-1 as part of the Priority Program SPP 2137 ``Skyrmionics''; via the projects C03 and C05 of the Collaborative Research Center SFB 1143 (project-id 247310070) at the TU Dresden; and the W\"{u}rzburg-Dresden Cluster of Excellence on Complexity and Topology in Quantum Matter -- \textit{ct.qmat} (EXC 2147, project-id 390858490). A.S.S. and B.E.Z.C. acknowledge support from the International Max Planck Research School for Chemistry and Physics of Quantum Materials (IMPRS-CPQM).


\begin{thebibliography}{}


\bibitem{Koshibae} W. Koshibae and N. Nagaosa, Nat. Commun. \textbf{7}, 10542 (2016).

\bibitem{Huang2017} S. Huang, C. Zhou, G. Chen, H. Shen, A. K. Schmid, K. Liu, and Y. Wu, Phys. Rev. B \textbf{96}, 144412 (2017).

\bibitem{Camosi} L. Camosi, N. Rougemaille, O. Fruchart, J. Vogel, and S. Rohart, Phys. Rev. B \textbf{97}, 134404 (2018).

\bibitem{Hoffmann} M. Hoffmann, B. Zimmermann, G. P. M{\"u}ller, D. Sch{\"u}rhoff, N. S. Kiselev, C. Melcher, and S. Bl{\"u}gel, Nat. Commun. \textbf{8}, 308 (2017).

\bibitem{Zhang} X. Zhang, J. Xia, Y. Zhou, X. Liu, H. Zhang, and M. Ezawa, Nat. Commun. \textbf{8}, 1717 (2017).

\bibitem{Nagaosa} N. Nagaosa and Y. Tokura, Nat. Nanotechnol. \textbf{8}, 899 (2013).

\bibitem{Muehlbauer} S. M{\"u}hlbauer, B. Binz, F. Jonietz, C. Pfleiderer, A. Rosch, A. Neubauer, R. Georgii, P. B{\"o}ni, Science \textbf{323}, 915 (2009).

\bibitem{Kindervater} J. Kindervater, T. Adams, A. Bauer, F. X. Haslbeck, A. Chacon, S. M{\"u}hlbauer, F. Jonietz, A. Neubauer, U. Gasser, G. Nagy, N. Martin, W. H{\"a}ussler, R. Georgii, M. Garst, and C. Pfleiderer, Phys. Rev. B \textbf{101}, 104406 (2020).

\bibitem{Yu_2} X. Z. Yu, N. Kanazawa, Y. Onose, K. Kimoto, W. Z. Zhang, S.  Ishiwata, Y. Matsui, and Y. Tokura, Nat. Mater. \textbf{10}, 106--109 (2011).

\bibitem{Wilhelm} H. Wilhelm, M. Baenitz, M. Schmidt, U. K. R{\"o}ssler, A. A. Leonov, and A. N. Bogdanov, Phys. Rev. Lett. \textbf{107}, 127203 (2011).

\bibitem{Milde2013} P. Milde, D. Köhler, J. Seidel, L.M. Eng, A. Bauer, A. Chacon, J. Kindervater, S. M\"{u}hlbauer, C. Pfleiderer, C. Sch\"{u}tte, and A. Rosch, Science \textbf{340}, 1076 (2013).

\bibitem{Moskvin} E. Moskvin, S. Grigoriev, V. Dyadkin, H. Eckerlebe, M. Baenitz, M. Schmidt, and H. Wilhelm, Phys. Rev. Lett. \textbf{110}, 077207 (2013).

\bibitem{Tokunaga} Y. Tokunaga, X. Z. Yu, J. S. White, H. M. R\o{}nnow, D. Morikawa, Y. Taguchi, and Y. Tokura, Nat. Commun. \textbf{6}, 7638 (2015).

\bibitem{Karube} K. Karube, J. S. White, N. Reynolds, J. L. Gavilano, H. Oike, A. Kikkawa, F. Kagawa, Y. Tokunaga, H. M. Ronnow, Y. Tokura, and Y. Taguchi, Nat. Mater. \textbf{15}, 1237 (2016).

\bibitem{Ukleev} V. Ukleev, Y. Yamasaki, D. Morikawa, K. Karube, K. Shibata, Y. Tokunaga, Y. Okamura, K. Amemiya, M. Valvidares, H. Nakao, Y. Taguchi, Y. Tokura, and T. Arima, Phys. Rev. B \textbf{99}, 144408 (2019).

\bibitem{Karube2} K. Karube, J. S. White, D. Morikawa, M. Bartkowiak, A. Kikkawa, Y. Tokunaga, T. Arima, H. M. Ronnow, Y. Tokura, and Y. Taguchi,
Phys. Rev. Materials \textbf{1}, 074405 (2017).

\bibitem{Karube3} K. Karube, J. S. White, D. Morikawa, C. D. Dewhurst, R. Cubitt, A. Kikkawa, X. Yu, Y. Tokunaga, T.-h. Arima, and H. M. Ronnow \textit{et al.}, Sci. Adv. \textbf{4}, eaar7043 (2018).

\bibitem{Seki12} S. Seki, X.  Z. Yu, S. Ishiwata, and Y. Tokura, Science \textbf{336}, 198 (2012).

\bibitem{Adams} T. Adams, A. Chacon, M. Wagner, A. Bauer, G. Brandl, B. Pedersen, H. Berger, P. Lemmens, and C. Pfleiderer, Phys. Rev. Lett. \textbf{108}, 237204 (2012).

\bibitem{Zhang2016} S. L. Zhang, A. Bauer, D. M. Burn, P. Milde, E. Neuber, L. M. Eng, H. Berger, C. Pfleiderer, G. van der Laan,
and T. Hesjedal, Nano Lett. \textbf{16}, 3285 (2016).

\bibitem{Milde2016} P. Milde, E. Neuber, A. Bauer, C. Pfleiderer, H. Berger, and L.M. Eng, Nano Lett. \textbf{16}, 5612 (2016).

\bibitem{Stefancic} A. Stefancic, S. H. Moody, T. J. Hicken, M. T. Birch, G. Balakrishnan, S. A. Barnett, M. Crisanti, J. S. O. Evans, S. J. R. Holt, K. J. A. Franke, P. D. Hatton, B. M. Huddart, M. R. Lees, F. L. Pratt, C. C. Tang, M. N. Wilson, F. Xiao, and T. Lancaster, Phys. Rev. Mater. 2, 111402(R) (2018).

\bibitem{Sukhanov} A. S. Sukhanov, P. Vir, A. S. Cameron, H. C. Wu, N. Martin, S. M{\"u}hlbauer, A. Heinemann, H. D. Yang, C. Felser, and D. S. Inosov, Phys. Rev. B \textbf{100}, 184408 (2019).

\bibitem{Kezsmarki} I. K\'{e}zsm\'{a}rki, S. Bord\'{a}cs, P. Milde, E. Neuber, L. M. Eng, J. S. White, H. M. R\o{}nnow, C. D. Dewhurst, M. Mochizuki, K. Yanai, H. Nakamura, D. Ehlers, V. Tsurkan, and A. Loidl, Nat. Mat. \textbf{14}, 1116 (2015).

\bibitem {Fujima}  Y. Fujima, N. Abe, Y. Tokunaga, and T. Arima, Phys. Rev. B \textbf{95}, 180410(R) (2017).

\bibitem{Bordacs} S. Bord\'{a}cs, A. Butykai, B. G. Szigeti, J. S. White, R. Cubitt, A. O. Leonov,  S. Widmann, D. Ehlers, H.-A. Krug von Nidda, V. Tsurkan, A. Loidl and I. K\'{e}zsm\'{a}rki, Sci. Rep. \textbf{7}, 7584 (2017). 

\bibitem{Nayak} A. K. Nayak, V. Kumar, T. Ma, P. Werner, E. Pippel, R. Sahoo, F. Damay, U. K. R{\"o}ssler, C. Felser, and S. S. P. Parkin, Nature (London) \textbf{548}, 561 (2017).

\bibitem{Saha} R. Saha, A. K. Srivastava, T. Ma, J. Jena, P. Werner, V. Kumar, C. Felser, and S. S. P. Parkin, Nat. Commun. \textbf{10}, 5305 (2019).

\bibitem{Jena} J. Jena, B. G{\"o}bel, T. Ma, V. Kumar, R. Saha, I. Mertig, C. Felser, and S. S. P. Parkin, Nat. Commun. \textbf{11}, 1115 (2020).

\bibitem{Peng} L. Peng, R. Takagi, W. Koshibae, K. Shibata, K. Nakajima,
T. Arima, N. Nagaosa, S. Seki, X. Yu, and Y. Tokura, Nat. Nanotechnol. \textbf{15}, 181 (2020).

\bibitem{Jena2} J. Jena, R. Stinshoff, R. Saha, A. K. Srivastava, T. Ma, H. Deniz, P. Werner, C. Felser, and S. S. P. Parkin, Nano Lett. \textbf{20}, 59--65 (2020).

\bibitem{Meshcheriakova } O. Meshcheriakova, S. Chadov, A. K. Nayak, U. K. R{\"o}ssler, J. K{\"u}bler, G. Andre, A. A. Tsirlin, J. Kiss, S. Hausdorf, A.
Kalache, W. Schnelle, M. Nicklas, and C. Felser, Phys. Rev. Lett. \textbf{113}, 087203 (2014).

\bibitem{Bogdanov } A. N. Bogdanov, U. K. R{\"o}ssler, M. Wolf, and K.-H. M{\"u}ller, Phys. Rev. B \textbf{66}, 214410 (2002).

\bibitem{Muehlbauer2} S. M{\"u}hlbauer, D. Honecker, E. A. Perigo, F. Bergner, S. Disch, A. Heinemann, S. Erokhin, D. Berkov, C. Leighton, M. R. Eskildsen, and A. Michels, Rev. Mod. Phys. \textbf{91}, 015004 (2019).

\bibitem{PA20} G. Chaboussant, S. Desert, P. Lavie, and A. Brulet, J. Phys.: Conf. Ser. \textbf{340}, 012002 (2012).

\bibitem{Supp} See Supplemental Material at [URL will be inserted by publisher] for additional information about the samples and complementary MFM measurements.

\bibitem{Vir1} P. Vir, J. Gayles, A. S. Sukhanov, N. Kumar, F. Damay, Y. Sun, J. K{\"u}bler, C. Shekhar, and C. Felser, Phys. Rev. B \textbf{99}, 140406(R) (2019).

\bibitem{Vir2} P. Vir, N. Kumar, H. Borrmann, B. Jamijansuren, G. Kreiner, C. Shekhar, and C. Felser, Chem. Mater. \textbf{31}, 5876 (2019).

\bibitem{Park} Park Systems Corp. KANC 15F, Gwanggyo-ro 109, Suwon 16229, Korea. 

\bibitem{Nanosensors} NANOSENSORS\texttrademark, Rue Jaquet-Droz 1, Case Postale 216, CH-2002 Neuchatel, Switzerland.

\bibitem{Omicron} Omicron NanoTechnology Gmbh, Taunusstein, Germany.

\bibitem{RHK} RHK Technology, Inc., 1050 East Maple Road, Troy, MI 48083 USA.

\bibitem{Gwyddion} D. Ne\v{a}as and P. Klapetek, Cent. Eur. J. Phys. \textbf{10}, 181 (2012).

\bibitem{Schmidt} P. W. Schmidt, in \textit{The Modern Aspects of Small-Angle
Scattering}, edited by H. Brumberger (Kluwer Academic,
Dordrecht, Netherlands, 1995), p. 1. 

\bibitem{Mildner} D. F. R. Mildner and P. L. Hall, J. Phys. D \textbf{19}, 1535 (1986).

\bibitem{Martin} J. E. Martin and A. J. Hurd, J. Appl. Crystallogr. \textbf{20}, 61 (1987).

\bibitem{Iashina} E. G. Iashina, E. V. Velichko, M. V. Filatov, W. G. Bouwman, C. P. Duif, A. Brulet, and S. V. Grigoriev, Phys. Rev. E \textbf{96}, 012411 (2017).

\bibitem{Kreyssig} A. Kreyssig, R. Prozorov, C. D. Dewhurst, P. C. Canfield, R. W. McCallum, and A. I. Goldman, Phys. Rev. Lett. \textbf{102}, 047204 (2009).

\bibitem{Pastushenkov1997} Yu.G. Pastushenkov, A. Forkl, H. Kronm\"{u}ller, J. Magn. Magn. Mater. \textbf{174}, 278 (1997).

\bibitem{Pastushenkov2010} Yu.G. Pastushenkov, Bulletin of the Russian Academy of Sciences: Physics \textbf{74}, 1423 (2010).

\bibitem{Hubert} A. Hubert and R. Sch\"{a}fer, \textit{Magnetic Domains -- The Analysis of Magnetic Microstructures}, Springer-Verlag Berlin Heidelberg (1998).

\bibitem{Kumar} V. Kumar, N. Kumar, M. Reehuis, J. Gayles, A. S. Sukhanov, A. Hoser, F. Damay, C. Shekhar, P. Adler, and C. Felser, Phys. Rev. B \textbf{101}, 014424 (2020).

\bibitem{Neubauer} A. Neubauer, C. Pfleiderer, B. Binz, A. Rosch, R. Ritz, P. G. Niklowitz, and P. B{\"o}ni, Phys. Rev. Lett. \textbf{102}, 186602 (2009).

\bibitem{Huang2012} S. X. Huang and C. L. Chien, Phys. Rev. Lett. \textbf{108}, 267201 (2012).

\bibitem{Gallagher} J. C. Gallagher, K. Y. Meng, J. T. Brangham, H. L. Wang, B. D. Esser, D. W. McComb, and F. Y. Yang, Phys. Rev. Lett. \textbf{118}, 027201 (2017).

\bibitem{Kanazawa} N. Kanazawa, Y. Onose, T. Arima, D. Okuyama, K. Ohoyama,
S. Wakimoto, K. Kakurai, S. Ishiwata, and Y. Tokura, Phys. Rev. Lett. \textbf{106}, 156603 (2011).

\bibitem{Yokouchi} T. Yokouchi, N. Kanazawa, A. Tsukazaki, Y. Kozuka, M.
Kawasaki, M. Ichikawa, F. Kagawa, and Y. Tokura, Phys. Rev. B \textbf{89}, 064416 (2014).

\bibitem{Li} Y. Li, N. Kanazawa, X. Z. Yu, A. Tsukazaki, M. Kawasaki, M. Ichikawa, X. F. Jin, F. Kagawa, and Y. Tokura, Phys. Rev. Lett. \textbf{110}, 117202 (2013).

\bibitem{Franz} C. Franz, F. Freimuth, A. Bauer, R. Ritz, C. Schnarr, C.
Duvinage, T. Adams, S. Bl{\"u}gel, A. Rosch, Y. Mokrousov, and C. Pfleiderer, Phys. Rev. Lett. \textbf{112}, 186601 (2014).

\bibitem{Ezawa} M. Ezawa, Phys. Rev. Lett. \textbf{105}, 197202 (2010).

\bibitem{Bernand} A. Bernand-Mantel, C. B. Muratov, and T. M. Simon, Phys. Rev. B \textbf{101}, 045416 (2020).

\bibitem{Yu} X. Yu, M. Mostovoy, Y. Tokunaga, W. Zhang, K. Kimoto, Y. Matsui, Y. Kaneko, N. Nagaosa, and Y. Tokura, Proc. Natl. Acad. Sci. U. S. A. \textbf{109}, 8856 (2012).

\bibitem{Takagi} R. Takagi, X. Z. Yu, J. S. White, K. Shibata, Y. Kaneko, G. Tatara, H. M. R\o{}nnow, Y. Tokura, and S. Seki, Phys. Rev. Lett. \textbf{120}, 037203 (2018).

\bibitem{Ma} T. Ma, A. K. Sharma, R. Saha, A. K. Srivastava, P. Werner, P. Vir, V. Kumar, C. Felser, and S. S. P. Parkin,  Adv. Mater. \textbf{n.a.}, 2002043 (2020).

\end{thebibliography}
\end{document}